# Emotion classification using EEG headset signals and Random Forest

Ricardo Vasquez
Universidad Internacional SEK
Quito, Ecuador
rfvasquez.mdat@uisek.edu.ec)

Diego Riofrío-Luzcando
Universidad Internacional SEK, Quito, Ecuador
Quito, Ecuador
diego.riofrio@uisek.edu.ec

Joe Carrion-Jumbo
(Universidad Internacional SEK, Quito, Ecuador
Quito, Ecuador
joe.carrion@uisek.edu.ec

Cesar Guevara
Centro de Investigación en Mecatrónica y Sistemas Interactivos - MIST
Universidad Indoamérica
Quito, Ecuador
DataLab
The Institute of Mathematical Sciences (ICMAT- CSIC) Madrid, Spain
cesarguevara@uti.edu.ec cesar.guevara@icmat.es

**Abstract:** Emotions are one of the important components of the human being, thus they are a valuable part of daily activities such as interaction with people, decision making and learning. For this reason, it is important to detect, recognize and understand emotions using computational systems to improve communication between people and machines, which would facilitate the ability of computers to understand the communication between humans. This study proposes the creation of a model that allows the classification of people's emotions based on their EEG signals, for which the brain-computer interface EMOTIV EPOC was used. This allowed the collection of electroencephalographic information from 50 people, all of whom were shown audiovisual resources that helped to provoke the desired mood. The information obtained was stored in a database for the generation of the model and the corresponding classification analysis. Random Forest model was created for emotion prediction (happiness, sadness and relaxation), based on the signals of any person. The results obtained were 97.21% accurate for happiness, 76% for relaxation and 76% for sadness. Finally, the model was used to generate a real-time emotion prediction algorithm; it captures the person's EEG signals, executes the generated algorithm and displays the result on the screen with the help of images representative of each emotion.

# 1 Introduction

In recent years, applications related to brain-computer interfaces have been one of the most popular topics in computer science since they allow communication between the human brain and a computer interface by capturing and interpreting EEG electroencephalographic signals.

This process has been evolving, avoiding being so invasive, but equally effective for the person, reducing the costs of medical equipment. It is worth mentioning that the use of this type of analysis is mostly limited to the health area and is not oriented to other branches such as education or the use of private individuals [Arboleda, 09]. Therefore a Random Forest model is proposed for the classification of EEG signals related to people's emotions.

# 2 State of the Art

This research was divided into three subtopics, emotion classification, EEG signal classification and classification using the Emotiv headset device. The main terms used in the search were "Emotion classification", "Random Forest classification", "Emotiv epoc", "EEG devices", "EEG classification", "Classification models", among others.

According to these search criteria we found several studies on emotion identification and classification using information from physiological responses such as facial features, voice recognition or body signals induced by music, videos, or other factors [Bailenson, 08]. For example, an investigation focuses on the analysis of reactions by capturing facial movements, the experiment was conducted with a sample of 50 people and consisted of showing them images to record their positive, negative, or neutral reactions. As a result, it was found that facial analysis systems can only detect external emotions and have great difficulty in detecting internal emotions, presenting very low ranges of accuracy [Magdin, 19].

Bailenson presented real-time automated models created with machine learning algorithms, using videotapes to elicit physiological responses to predict sadness or amusement. For this, algorithms based on points extracted from subjects' faces were created, as well as their physiological responses such as cardiovascular activity, somatic activity, and electrodermal response. The results showed better performance for ratings of amusement than sadness [Bailenson, 08].

Ververidis proposes the design of a tool that automatically classifies five emotional expressions: anger, happiness, neutrality, sadness and surprise. To do that, they relied on a speech analysis software tool developed in MATLAB which records speech and displays it on screen as a waveform. The result is a correct classification rate of 51.6% ± 3% with a 95% confidence interval for all five emotions, while a random classification would give a correct classification rate of 20% [Ververidis, 04].

## 2.1 Classification of EEG signals using different models

There are several articles that talk about EEG signal classification using brain-computer interfaces, among the main ones we can mention those related to home automation and emotional mood recognition [Bos, 06; Ghodake, 16; Li, 19; Lin, 10; Nugraha, 16; Roy, 20; Shukla, 18; Virdi, 17]. For example, Roy uses the 14-channel Emotiv Epoc+ device to obtain EEG signals from individuals to monitor a bulb. They achieved this by sorting the EEG signals with neural networks to be sent to a circuit that interpreted the action and executed it [Roy, 20].

Lin conducted a study in which machine learning algorithms were applied to categorize EEG signals, pertaining to the mood of people, these states were happiness, anger, sadness and pleasure. The study had a population of 26 people and used the support vector machine classification algorithm which returned an average accuracy of 82.29% ± 3.06% [Lin, 10].

While Li aimed to present an emotion recognition system using classification and pattern recognition techniques. Videos were used as stimuli for each emotion and the emotions studied were happiness, relaxation and sadness. For the study, the relevance vector machine (RVM) classification algorithm was used on a population of 5 people. The result was an accuracy in happiness of 70.70%, relaxation 73.30% and sadness 72.30% [Li, 19].

Meanwhile Nugraha lead a study to determine whether a person is fatigued when driving, using EEG signals from 30 people as input data. They used the kNN and SVM algorithms to classify the data. With the kNN algorithm they obtained an accuracy of 96% and with the SVM algorithm they achieved an accuracy of 81% [Nugraha, 16].

**2.2    Classification of EEG signals using Random Forest models**

On the other hand, several papers deal with the classification of EEG signals with Random Forest models for various applications, among the most important of which are medical applications [Chen, 14; Donos, 15; Edla, 18; Shen, 07]. For example, Chen performed classification on EEG signals to detect brain disorders in newborns [Chen, 14]. While Donos developed an early seizure detection algorithm based on of intracranial EEG signals using Random Forest [Donos, 15].

On the other hand, Edla performed EEG data classification for the analysis of human mental state using Random Forest where the analyzed mental states were concentration and meditation. The data collection was performed using the Neurosky Mindwave Mobile device of 8 values and had a sample of 40 subjects obtaining as a result a correct prediction of 75% [Edla, 18].

**2.3    Classification of EEG signals using Emotiv headset**

In addition, several investigations on EEG signal classification using Emotiv headset device for data collection [Liu, 11; Nugraha, 16; Yu, 16]. For example, Liu proposed a real-time emotion classification algorithm from EEG signals to analyze the state of fear, frustration, sadness, happiness, and contentment. These emotions were elicited with the help of digital sounds, playback of musical pieces and questionnaires [Liu, 11].

The study by Liu collected data from 22 individuals using the 14-channel Emotiv headset device of which three were used for the study AF3, F4 and FC6. The result was a prototype real-time emotion recognition system that allows the visualization of emotions in the form of facial expressions of personalized avatars in 3D environments [Liu, 11].

## 3   Method

The process to achieve a classification of emotions is divided generally in six steps: Generation of emotions in people, reading of EEG signals with the device, capturing the signals sent through a tailor-made software, storage into a database, analysis of stored information, presentation of study results and display of alive results (Figure 1).

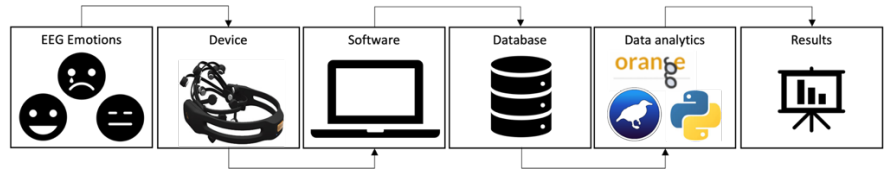

*Figure 1 - General Diagram of the Process.*

With these steps, an analysis was conducted that obtained people's EEG signals as input data, and the mood of these signals was predicted by means of a classification process. In order to reach this goal, the following processes were implemented: Data collection, analysis, validation and classification in real time.

### 3.1   Data collection

Data collection involved acquisition, processing and storage of the necessary data for generation of the classification model. These data were made up by EEG signals from different people corresponding to three emotions: happiness, sadness and relaxation. These signals were extracted with the EMOTIV EPOC device, which is an interface for interaction between the brain and a 14-channel computer, which allowed communication between the brain and the data collection system.

To gather the necessary information, an experiment was conducted with a sample of 50 volunteers of ages ranging from 17 to 71; 27 of them were men and 23 were women. The tasks detailed below were applied to each of the participants:

- Preparation of the EMOTIV EPOC device. This entailed verification that each sponge was hydrated with saline solution.
- Placement of the device on each participant
- Start of the data collection software.
    - System initialization by entering the participant's code and the action that must be stored.
    - Verification of sensors connection status by color codes black, red, orange and green.
- System calibration. At this point, information, such as age, gender, civil status, level of education, among others, was collected for each person so that all the sensors may have an excellent signal and to store the general information of each person in the database.
- Finally, the participants' mood data were collected by using audiovisual resources. Once the video starts, the recording of data into the system runs for five minutes, during which time, approximately 2000 records per minute are stored, which means that an average of 8000 records per action was recorded per each person (Figure 3).

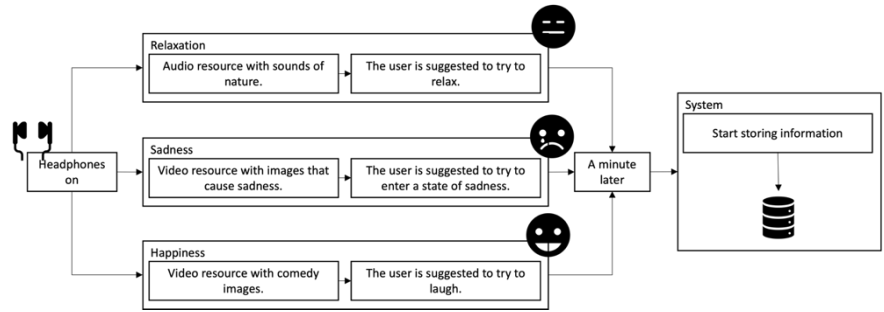

*Figure 2 - Data Collection Process*

As a result of this experiment, 1.106,752 data sets were obtained corresponding to the 3 actions of 50 people; these data were stored for subsequent tabulation and analysis. Once the data were collected, cleanup and model design processes were implemented, which allowed data classification.

The WEKA[1] tool was used for cleanup in order to identify and remove outlier data. The InterquartileRange filter was applied to mark these records, with the factor to determine extreme value thresholds as six and the factor for atypical thresholds as three. The detail of outlier data may be observed in Table 1.

| Class | Collected | Outliers | Total |
| --- | --- | --- | --- |
| RELAXED | 395281 | 41009 | 354272 |
| HAPPY | 355852 | 13442 | 342410 |
| SAD | 355619 | 34043 | 321576 |
| TOTAL | 1106752 | 88494 | 1018258 |

*Table 1 Outlier Data.*

Once outlier data were marked and excluded, an analysis was conducted between five classification models by using the Orange[2] program for the purpose of selecting the one offering the best results for the experiment. For the selection of these models, the number of variables, data dispersion and the amount of results the model could produce (happiness, sadness and relaxation) were taken into account. That is how the decision-tree models SVM, Random Forest, kNN and Neural Networks were selected.

This resulted in the information of Table 2, which showed the precision value and the area under curve (AUC) for each class. These data were used to establish that the best precision level was produced with the Random Forest model, with 97.20 % of correct classification.

---

[1] http://www.cs.waikato.ac.nz/ml/weka/
[2] http://orange.biolab.si/

| Color | Accuracy | AUC Happiness | AUC Relaxed | AUC Sad |
|---|---|---|---|---|
| kNN | 0.960 | 0.994 | 0.996 | 0.996 |
| Neural Network | 0.879 | 0.959 | 0.979 | 0.967 |
| Random Forest | 0.972 | 0.997 | 0.998 | 0.998 |
| SVM | 0.412 | 0.344 | 0.437 | 0.422 |
| Tree | 0.878 | 0.893 | 0.914 | 0.908 |

*Table 2 Accuracy and AUC Data for model design*

Finally, with this result, the classification process was generated by using Python[3] and the Scikit-learn library for generation of the Random Forest model and the data prediction.

The first step in the process was determining the optimal value of the model branches; in order to identify this value, a crossed validation process was implemented by changing this parameter from one until the accuracy difference between the two last models analyzed comes close to zero. As shown in Figure 3, upon reaching 25 branches of the confidence percentage of the prediction, a point is reached where the curve begins to flatten. This value was used to generate the Python model that will be used in the experiment.

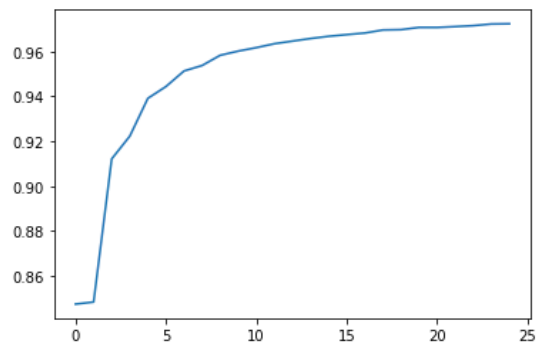

*Figure 3 Branch Number Analysis*

### 3.2 Validation

The crossed validation process was automated in Python and involved uploading the generated models, taking data from the prediction table, generating the prediction value for each record to finally store it in the database.

The following steps were taken for the validation:
- Generation of the entire model: For this, 30 test models of crossed validation were performed, beginning by an analysis that took into account 90 % of random data for model generation and 10 % of random data for validation generation. To obtain an amount of 30 models, a normality test of success percentages was conducted, resulting in a p-value of 0.6982; thus, no more models had to be added to the analysis.
- Graphical representation of data: It involved using the EEG signal data for a graphical representation, as shown in Figure 6. This graph shows similarity between relaxation and sadness; for this reason, a second model was performed by grouping emotions in pairs.
- Generation of grouped models: This was done by generating models grouped as joy-sadness, joy-relaxation and relaxation-sadness. For this validation, 250,000 records were randomly taken; 90 % of these were used in the generation of each model and 10 %, for crossed validation. This study was conducted to determine whether relaxation and sadness are confused one with the other.

---
[3] https://www.python.org/doc/

### 3.3 Data Classification in Real Time

This was done by generating a prototype that classifies signals in real time, which was developed with Microsof.Net C# and Python. This process was implemented with the following actions:
- The device was placed on the person and then the audiovisual was started.
- There was a waiting time of approximately 30 seconds for calibration before the data collection started.
- The data collection, the reading and the interpretation of signals through the system developed in C# was performed for ten seconds.
- Interpreted signals were sent to Python scripts, which uploaded the classification models and responded with the respective prediction again to C#.
- Finally, the system visually displayed the result on the screen in real time.
- The collection, interpretation and response processes are repeated every ten seconds.
- The length of the experiment was five minutes for each emotion, from the first step to the last.

## 4 Results

Results obtained by application of the Random Forest model to the data collected from a population of 50 people in total, ranging between 17 and 71 years of age (Table 3).

| Item | Breakdown |
|---|---|
| Total Population | 50 |
| Male Population | 27 |
| Female Population | 23 |
| Age Range | 17-71 years old |
| Nationality | Ecuadorean |

*Table 3 Population Breakdown*

Three experiments were conducted; the first one by generating a model based on all the data; the second one by grouping data by pairs: happiness-sadness, happiness-relaxation and relaxation-sadness; and the third one, by using the model with all the data for a real time prediction. The information used in the experiments is shown in Table 4.

| Breakdown | Quantity |
|---|---|
| Total collected | 1,106,752 |
| Outlier Data | 88,494 |
| Working Data | 1,018,258 |
| Relaxed Data | 354,272 |
| Happy Data | 342,410 |
| Sad Data | 321,576 |
| Experiment 1 Model Generation | 900,000 |
| Data for predictions 1 | 100,000 |
| Experiment 2 Model Generation | 225,000 |
| Data for predictions 2 | 25,000 |

*Table 4 General Data of Experiment*

## 4.1 Crossed validation for the whole model

For the first validation, 30 models generated with all the emotions were created, where each of them contained 900,000 records for creation and 100,000 records to be validated. The average precision result of the experiments was 97.69 %.

| N° | Correct | Incorrect | Accuracy | N° | Correct | Incorrect | Accuracy |
|---|---|---|---|---|---|---|---|
| 1 | 97704 | 2296 | 97.70% | 16 | 97676 | 2324 | 97.68% |
| 2 | 97613 | 2387 | 97.61% | 17 | 97760 | 2240 | 97.76% |
| 3 | 97569 | 2431 | 97.57% | 18 | 97626 | 2374 | 97.63% |
| 4 | 97772 | 2228 | 97.77% | 19 | 97718 | 2282 | 97.72% |
| 5 | 97741 | 2259 | 97.74% | 20 | 97643 | 2357 | 97.64% |
| 6 | 97686 | 2314 | 97.69% | 21 | 97760 | 2240 | 97.76% |
| 7 | 97654 | 2346 | 97.65% | 22 | 97625 | 2375 | 97.63% |
| 8 | 97692 | 2308 | 97.69% | 23 | 97616 | 2384 | 97.62% |
| 9 | 97695 | 2305 | 97.70% | 24 | 97658 | 2342 | 97.66% |
| 10 | 97644 | 2356 | 97.64% | 25 | 97778 | 2222 | 97.78% |
| 11 | 97697 | 2303 | 97.70% | 26 | 97827 | 2173 | 97.83% |
| 12 | 97681 | 2319 | 97.68% | 27 | 97640 | 2360 | 97.64% |
| 13 | 97656 | 2344 | 97.66% | 28 | 97663 | 2337 | 97.66% |
| 14 | 97716 | 2284 | 97.72% | 29 | 97720 | 2280 | 97.72% |
| 15 | 97782 | 2218 | 97.78% | 30 | 97641 | 2359 | 97.64% |

*Table 5 Experiment 1 Results.*

As can be seen in Table 5 there is a very high accuracy in all the results, however, in Figure 6 it could be observed that there is not much difference between relaxation and sadness, so it was necessary to perform a validation by class.

### 4.2  Cross-validation of clustered models

For this validation, 30 models were performed for each mood state, resulting in 97.21% happy, 76.53% relaxed and 76.04% sad. Table 6 shows the average confusion matrix of the 30 validations performed per class.

| Class | Happy | Relaxed | Sad | Accuracy |
| --- | --- | --- | --- | --- |
| HAPPY | 24303 | 206 | 491 | 97.21% |
| RELAXED | 5000 | 17610 | 400 | 76.53% |
| SAD | 5766 | 224 | 19011 | 76.04% |

*Table 6 Validation mix-up by class Matrix*

By means of these data it can be said that happiness is the class with the highest prediction ease while sadness and relaxation have a higher prediction difficulty. This may be due to the fact that the relaxation and sadness signals are similar according to what can be observed in Figure 5.

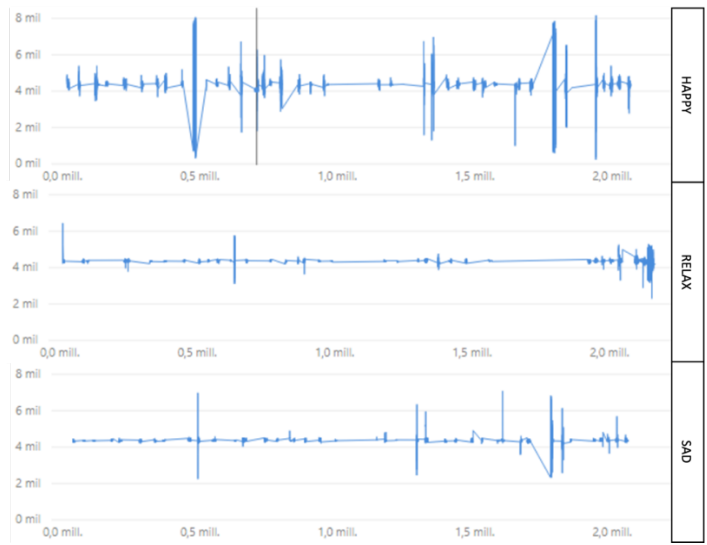

*Figure 5 Collected Signals.*

## 4.3 Real-time data validation

Real-time validation was performed with two volunteers who were fitted with the device and, with the help of audiovisual resources, reached the mood states (happy, relaxed, or sad).

The result was presented in real time on the computer monitor showing accuracy values between 35% and 90% (Table 7). The screen observed by the volunteers is shown in Figure 6. While Table 7 shows the data from real-time observations taken from one of the volunteers for five minutes with a time lag between each observation of 20 seconds.

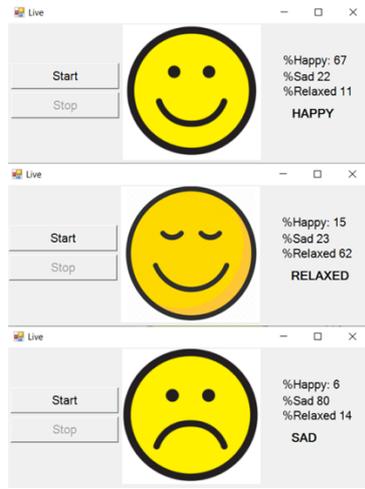

| Observation | Class | Accuracy | Time |
|---|---|---|---|
| 1 | HAPPINESS | 39% | 0:20 |
| 3 | HAPPINESS | 35% | 0:40 |
| 4 | HAPPINESS | 45% | 1:00 |
| 5 | HAPPINESS | 62% | 1:20 |
| 6 | HAPPINESS | 52% | 1:40 |
| 7 | HAPPINESS | 71% | 2:00 |
| 8 | HAPPINESS | 64% | 2:20 |
| 9 | HAPPINESS | 75% | 2:40 |
| 10 | HAPPINESS | 78% | 3:00 |
| 11 | HAPPINESS | 85% | 3:20 |
| 12 | HAPPINESS | 88% | 3:40 |
| 13 | HAPPINESS | 90% | 4:00 |
| 14 | HAPPINESS | 59% | 4:20 |
| 15 | HAPPINESS | 86% | 4:40 |
| 16 | HAPPINESS | 78% | 5:00 |

*Figure 6 Real-time System*          *Table 7 Real-time Observatio*

Finally, the data generated were stored and plotted online for comparison with the signals used in the model (Figure 7).

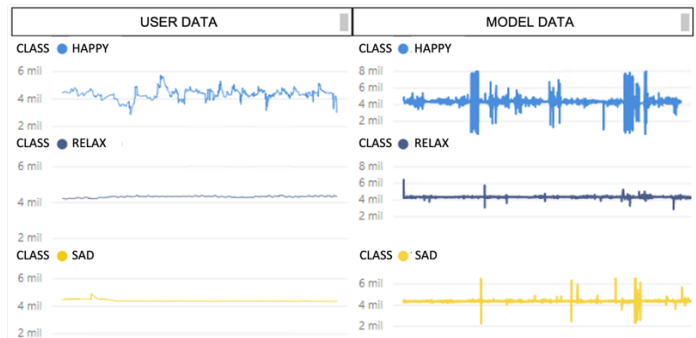

*Figure 1 Real-time signal results.*

As can be seen in Figure 7, the graphs have a similar pattern between those generated by a person in real time and those stored in the model. In the same way, it can be determined that there is much similarity between the relaxation and sadness graphs while happiness is quite different from the other two. To verify this similarity, we proceeded to calculate the overall variance of the model per sensor and the variance of the signals obtained in real time in Table 8.

| Model data variance | | | | | | | |
|---|---|---|---|---|---|---|---|
| CLASS | AF3 | F7 | F3 | FC5 | T7 | P7 | O1 |
| Happiness | 226269 | 367334 | 316039 | 243316 | 374687 | 201619 | 233778 |
| Relaxed | 14656 | 20666 | 11107 | 5281 | 21369 | 8124 | 16317 |
| Sadness | 12786 | 22248 | 18374 | 12880 | 28033 | 13029 | 16854 |
| User data variance | | | | | | | |
| CLASS | AF3 | F7 | F3 | FC5 | T7 | P7 | O1 |
| Happiness | 170481 | 252178 | 172178 | 67411 | 200134 | 93142 | 191941 |
| Relaxed | 2457 | 2306 | 2455 | 3255 | 2535 | 1446 | 1969 |
| Sadness | 8312 | 4059 | 7498 | 1386 | 5990 | 3852 | 6427 |
| Model data variance | | | | | | | |
| CLASS | O2 | P8 | T8 | FC6 | F4 | F8 | AF4 |
| Happiness | 326195 | 516709 | 396714 | 399771 | 238290 | 428539 | 388021 |
| Relaxed | 10495 | 65718 | 21795 | 38527 | 9494 | 32443 | 22022 |
| **Sadness** | 20878 | 45938 | 27366 | 30777 | 12233 | 31046 | 23377 |
| User data variance | | | | | | | |
| CLASS | O2 | P8 | T8 | FC6 | F4 | F8 | AF4 |
| Happiness | 161406 | 244196 | 164116 | 144113 | 136779 | 161727 | 157409 |
| Relaxed | 2464 | 3056 | 3121 | 2891 | 3128 | 2818 | 2662 |
| Sadness | 6649 | 7525 | 7501 | 8445 | 6606 | 7989 | 6823 |

*Table 8 Variance Comparison*

Figure 8 shows the graph of the variances, both in the figure and in the data, we can see that the variance in the case of happiness has high values of six low figures while for relaxation and sadness low values are obtained in the range of up to five low figures.

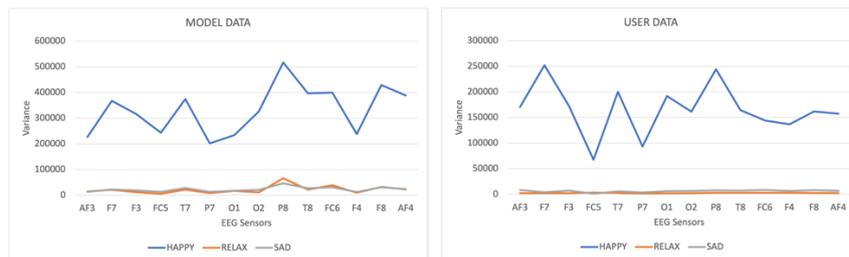

*Figure 2 Variance Comparison*

## 5 Conclusions and Future Work

In this study, a process of classification of the electroencephalographic signals of people using the Random Forest model was proposed, achieving the recognition of three emotions: happiness, sadness and relaxation. For this, a process was designed that allowed the collection of EEG information in a fast and simple way, this was achieved thanks to the development of an automated system of acquisition and storage of the information, allowing the process to take only 15 to 20 minutes per person.

The first analysis carried out with all the data of the 50 people showed an accuracy of 97.06%, while the second analysis showed a percentage of accuracy of 97.21% for happiness and 76% for relaxation and sadness, which allows us to conclude that relaxation and sadness have a lower percentage of accuracy, this could be because the signals of these emotions are similar to each other, which could have been caused by the bias towards happiness caused by the greater amount of data of that kind, which were obtained after cleaning the information.

In the third analysis, it was possible to develop a prototype that classifies the electroencephalographic signals of a person in real time and allows visualizing the mood of the person at that moment by means of graphs. According to the results of this analysis, it was possible to conclude that it was possible to determine in real time the mood of a person with an average accuracy rate of 67%. It could be observed that as time goes by, the volunteer concentrates more on the experiment, obtaining better results.

In general, it was possible to implement the recognition of the three emotions; however, the classification of sadness and relaxation could be improved both in the models and in the real-time recognition system. This process could be achieved by improving the process of taking signals with the help of health professionals. In addition, models with a greater number of emotions, such as meditation, anger, anguish, etc., could be created in the same way.